# The First 50 Years of Software Reliability Engineering: A History of SRE with First Person Accounts

**James J. Cusick, PMP**
New York, NY
j.cusick@computer.org

**Abstract — Software Reliability has just passed the 50-year milestone as a technical discipline along with Software Engineering. This paper traces the roots of Software Reliability Engineering (SRE) from its pre-software history to the beginnings of the field with the first software reliability model in 1967 through its maturation in the 1980s to the current challenges in proving application reliability on smartphones and in other areas. This history began as a thesis proposal for a History of Science research program and includes multiple previously unpublished interviews with founders of the field. The project evolved to also provide a survey of the development of SRE from notable prior histories and from citations of new work in the field including reliability applications to Agile Methods. This history concludes at the modern-day providing bookends in the theory, models, literature, and practice of Software Reliability Engineering from 1968 to 2018 and pointing towards new opportunities to deepen and broaden the field.**

## I. Introduction

Fifty years is a natural and major milestone. From a historical perspective we can establish that the practice of software reliability emerged in 1968 and since we have just concluded the year 2018 it is well to mark this anniversary as this paper attempts to do. While no single event may claim to represent the absolute beginning of the field and much work was being done in parallel, the 50 years between 1968 and 2018 does provide a convenient interval to review.

The key milestones marking the initiation of the field of Software Reliability Engineering include:

- Hudson's first software reliability model - 1967
- The NATO Software Engineering conference - 1968
- Shooman's text on probability – 1968
- Jelinkski and Miranda's reliability model - 1971

These accomplishments and those of other individuals who developed reliability methods and techniques began much of their work in the period beginning just around 50 years ago. In this paper key definitions are provided around reliability, Software Engineering, software reliability, and Software Reliability Engineering (SRE). The first steps taken towards more reliable software are retraced by those who were there through their writings and in previously unpublished interviews. In addition, this history is carried up to the present day since the last major treatment on this subject was published 35 years ago (Shooman, 1984). Finally, a tour of related "modern" methods meant to improve Software Engineering and reliability such as Agile methods are discussed.

### A. *Treatments of Reliability History*

The first major history of reliability was conducted by Marty Shooman in his 1984 article (Shooman, 1984). This paper provided a comprehensive review of applications of reliability in software up until that time, however, it is now 35 years old and some new information can be provided including documenting the last few decades of development in the field. While others such as Musa (1987) and Lyu (1996) touched on the history of the field there have been few treatments as comprehensive as Shooman's. It is hoped this paper can retrace that history, add some new details, and extend it to the present day as Software Reliability Engineering remains a critical discipline within Software Engineering.

Other notable early histories of computing do not touch on software or reliability directly. For example, Goldstine's comprehensive treatment of computer history focuses on computers themselves from Pascal to Babbage to van Neumann (Goldstine, 1972). However, he says nothing about reliability. Instead he focuses on the applications of the time especially mathematical, scientific, census, and ballistic problems as well as the speed of computation affecting the solutions to these problems.

Even break out business books of this early software period such as Phil Crosby's "Quality is Free" mentions reliability only in passing (Crosby, 1979). He focuses on the "design time" aspects of improving quality and introduces a maturity matrix to prompt management to think about quality in a more improvement-oriented manner across the board which later influenced the rise of the CMM by Humphrey (1989).

### *B. Structure of This Paper*

This paper begins with a summary of core concepts in reliability as practiced in hardware as those methods were later converted to use in software. This is followed by a recap of the proceedings of the 1968 NATO Conference on Software Engineering. In particular, we will focus on the discussions related to reliability at that dawn of the engineering treatment of software. This is followed by a definition and review of Software Engineering and Software Measurement in order to put software reliability into a proper context. After establishing the development of fundamental Software Reliability approaches, we will review the work and viewpoints of several early researchers including Glen Meyers and Marty Shooman. Next, we will present previously unpublished interviews from Norm Shneidewind and John Musa who worked both independently and jointly on pioneering reliability research and practice in the early 1970s. Following this we will review the progress of the 1990s and 2000s and finally look at the various current methods in Software Engineering which can positively impact software reliability. Finally, a single page timeline of this entire history is appended for reference at the conclusion of this paper.

### *C. Software Reliability in Context*

Software is now deeply embedded in the infrastructure of society like never before. In the contemporary world of technology software plays a vital part in our everyday life. We rely on software for traffic lights, airplane guidance, stock trades, and medical devices. In life critical systems software plays an even more significant role. Airplanes cannot fly without software, modern cars cannot stop without software, and emergency calls cannot be placed without software. Today getting a reliable software version is a science and engineering discipline that is well documented and proven.

To create this software there are innumerous methods at the disposal of the software professional – but reliability is always a factor and needs to be understood in advance of use and during use. In the earliest days of software development, it was unknown in advance what the reliability level of a software product or release would be (Naur, 1968). Furthermore, the dynamic failure rates in production usage could not be predicted due to an unknow level of latent defects in the software.

These and related problems were those which early researchers and practitioners investigating software reliability began looking at in the late 1960s and early 1970s. The first reliability model was provided by Hudson in 1967. This was quickly followed by Shooman's landmark book on probabilistic reliability published in 1968 which provided much of the theoretical and practical convergence of reliability methods as used in hardware reliability for later use in software reliability. The first published software reliability models came from Jelinkski and Miranda in 1971 & 1972 (Jelinkski, 1971; Jelinski, 1972) and independently at the same time from Shneidewind (Shneidewind, 2006). By the late 1970s most of the theory supporting a predictive reliability practice had been created by many of the researches discussed in this paper (Musa, 2006). This work has resulted in the ability of the Software Engineering field to develop, deploy, and operate systems of high reliability to manage not only routine applications but especially those with life-critical requirements. A routine practice of the field was eventually built up in the 1980s and 1990s (Lyu, 1996).

## II. BACKGROUND

This work has as a central goal to discover and document how the science and engineering discipline of Software Reliability Engineering (SRE) emerged, was formalized, and popularized. The perspective taken here emanates from the field of History of Science. Thus, of interest are the precursor events, influences, individuals, and significant developments, methods, and technologies leading to the formation and evolution of this technical field. This exploration will touch on the scientific aspects of SREs development but primarily will focus on the practice of this applied engineering method. This investigation will lead us to the roots of current reliability methods which lie primarily in hardware reliability methods, statistics, and probability theory beginning in the early 19th century. This work will explore how these areas developed and led to the emergence of a general understanding of software reliability as a characteristic of software quality and driven by non-functional requirements. as well as what novel inventions were required to accomplish this[1].

We will also review how the achievement of software reliability was codified into a practice called SRE during the 1970s and 1980s, including what theoretical problems were encountered and how the practice was routinized. The practice of SRE has been well documented by many authors referenced here but the origins of the field have been less fully defined and can benefit from additional detail. The manner in which this engineering practice became widespread will also be explored and documented. The effect of technical evangelism, the use of technical community outreach, and the culture around SRE will all be covered. In the end a complete picture will emerge of how predecessor technological ideas and contemporary engineering problems along with the inspiration of a group of scientists and engineers provided a new and well-structured software engineering capability in general and widespread use today.

Formally quantified reliability measures assume a higher level of maturity in process, engineering, and methodology. While the methods of reliability engineering do not require this sophistication, it is common for those who develop and deploy these methods to approach their work in a more structured manner. Thus, one will see most users of reliability engineering falling into the process-oriented team category but not in all cases. As will be shown, even in some Agile process teams the same formal reliability methods are used.

[1] The author was introduced to SRE in 1992 as a practitioner roughly at the halfway mark between the founding of the field in 1968 and today 2018 and has continued to use SRE methods on and off since then as well as to publish experience reports on this work. This involvement in the use of SRE provides the research in this paper around this history of professional interest from an engineer's standpoint and not only as a historian.

This paper discusses what is meant by software engineering in order to introduce Software Reliability Engineering and reviews key literature on the establishment of the field including the research of McLinn who also traced the history of reliability overall including a brief discussion of software reliability. Of special note are edited interview transcripts from two leaders in the field, John Musa and Norm Schneidewind, included below, who were both contacted regarding this study and provided input and suggestions on this paper to help ensure relevance to the field of Software Reliability Engineering. These first-person interviews are unique to this history and provide an eyewitness touch to the research as well as add some previously undocumented details.

## III. Software Reliability's Roots in Reliability Engineering

Software Reliability Engineering is a field that developed from roots within the reliability disciplines of structural, electrical, and hardware engineering (Shooman, 1984, Schneidewind, 2006). These disciplines had well established concepts around reliability and methods for determining its values. Within structural engineering concepts of stress and failure were used to calculate reliability. In hardware engineering, from electronics to electrical engineering and integrated circuit design, concepts of reliability were well established. The actual origins of Reliability Engineering can be found in the early 19th century industrial world:

> *"From its modest beginning in 1816 - the word reliability was first coined by Samuel T. Coleridge - reliability grew into an omnipresent attribute with qualitative and quantitative connotations that pervades every aspect of our present day technologically intensive world."* (Saleh, 2006)

The development of reliability engineering slackened until the mid-20th century and was revived with the evolution of statistics and the demands of mass production (Saleh, 2006). However, refinement of these ideas by Dr. Walter A. Shewhart of Bell Labs, who began formulating the concepts of statistical quality control changed the path of reliability research and application. His first solution was presented in 1924 as a basic statistical control chart which is a core aspect of understanding the reliability of processes even today (Shewhart, 1931). His concepts would go on to provide the foundations of much of modern manufacturing quality methods.

Leading up to World War II, "reliability as a word came to mean dependability or repeatability" (McLinn, 2010). Today's usage of the term reliability became more recognizable as the U.S. military evolved its application of such methods in the 1940s leading to many of the present connotations. Originally the term meant that a product would operate when expected (McLinn, 2010).

Especially in the 1950s the new understanding of reliability grew out of "the catalyst that accelerated the coming of this new discipline [which] was the (unreliability of the) vacuum tube" (Saleh, 2006). Other key milestones included work by Weibull on statistical models, the formation of the IEEE Reliability Society in 1948, and the creation by the DoD of a group called AGREE to standardize reliability approaches (Tan, 2017). Each of these events and developments set the stage for the investigation and development of software reliability practice in the next decade as simultaneously the field of Software Engineering was launched.

## IV. Reliability at the 1968 NATO Conference

### A. The Conference in Overview

The emergence of Software Reliability Engineering can be said to have one of its roots in the first conference on Software Engineering. It was at the NATO Conference on Software Engineering in 1968 that the term Software Engineering was first formally used (Pfleeger, 1998) and at that conference software reliability received attention and even its own focus session (Naur, 1968).

The topics covered in the NATO conferences on Software Engineering in 1968 and 1969 attempted to provoke research and investigation in how the computing field might be able to develop solutions in a more predictable basis as in the engineering and civil engineering fields relying on proven methods and practice and delivering expected results (van Vliet, 1993). Within this also lay the emergence of the practice of Software Reliability Engineering as one answer to the challenge.

About 50 invited attendees participated including such notables as Dijkstra of the Technologic University, Bauer of the Institut der Technischen Hochschule, McClure of Bell Labs, Nash of IBM Labs UK, etc. (Naur, 1968). The proceedings are certainly worth a read and by way of context establish that in Europe at the time only 10,000 computers were operational, but they grew at 25-50% per year. Discussions were dominated by references to IBM OS/360, Multics, OS and compiler development, large-scale systems such as SABRE, and the advent of time-sharing systems. Surprisingly, many modern ideas including the discussion of iterative development alongside consideration of some intractable problems such as software quality (and by extension reliability) and effort estimation were also covered at length (Naur, 1968).

### B. NATO Conference and Reliability

Discussions within the conference mentioned the problems of achieving sufficient reliability in the data systems at the time which were becoming increasingly integrated into the central activities of modern society. Examples of slipped schedules, extensive rewriting, much lost effort, large numbers of bugs, and inflexible and unwieldy products were described by many participants. They noted that products developed under these conditions were not likely to be brought to a satisfactory state of reliability or that they could be maintained and modified.

In his comments on reliability Dijkstra (Naur, 1968) explained that he was …

> *"...convinced that the quality of the product can never be established afterwards. Whether the correctness of a piece of software can be guaranteed or not depends greatly on the structure of the thing made. This means that the ability to convince users, or yourself, that the product is good, is closely intertwined with the design process itself."* (Naur, 1968)

Other speakers agreed and several commented that reliability really is a design issue. Thus, we can see even at this early stage in Software Engineering that the non-functional requirement of reliability was quite central and was both a major factor in product delivery and that it emerged from design and could not necessarily be "tested into" the product.

However, we do come to testing in the development and delivery process in the proceedings. One of the closest forerunners of the concepts we see in Software Reliability Engineering is that of d'Agapeyeff who describes how an "increase [in] run time checks [can] therefore [enhance] program reliability" (Naur, 1968). This foreshadows the concept of the Operational Profile driven probabilistic testing methods to come (Musa, 1993).

### *C. Testing and Reliability at NATO Conference*

The conference also discussed testing in all its phases as understood at that time. The perspective presented by Opler was that…

> *"...a test plan must be developed considering all elements of the written specification (hardware, programming language, system facilities, documentation, performance, reliability, etc.) and describing steps to validate compliance of the final programming system."* (Naur, 1968)

Furthermore, Llewelyn and Wickens presented a perspective on testing that stated that a testing "is one of the foundations of all scientific enterprise" (Naur, 1968; Cusick, 2018). They also stated that the methodology for testing at that time called for availability and acceptance testing after the fact as the state of the art at that time.
Unfortunately, they noted that at the time "…the present situation is that a customer has to purchase his software almost as an act of faith in the supplier …". However, they wanted to see testing emerge into a discipline on an equal footing with the rest of computing and Software Engineering. What they called for was a method whereby …

> *"...one can measure [software's] performance experimentally and see if the results are in accord with the specification; the number and sophistication of the experiments can be varied to provide the degree of confidence required of the results."* (Naur, 1968)

Again, this kind of construct is fully aligned with what would become the defined field of SRE where sample sizes of test runs were drawn by an infinite pool through probability driven Operational Profiles and reliability figures are proven within predefined confidence levels. Thus, the thinking of Llwelyn and Wickens was directionally aligned with the SRE methodologists who were developing these ideas practically at the same time as this conference (see Hudson, 1967) and then followed just a few years later but with more detailed models and approaches.

### *D. Limits of the Day*

In reading the NATO conference proceedings one also sees clearly the limits of the state of the art at that time and the fact that the attendees were well aware of these limits. In fact, it is striking that much of the folklore and common sayings one hears today in the industry 50 years hence are clearly stated by those participants. As an example, here is Dijkstra on using testing to understand software correctness: "Testing is a very inefficient way of convincing oneself of the correctness of a program" (Naur, 1968). At the time correctness proofs coming from a mathematical perspective held sway over real-time and operational reliability testing.

A further extension to this by Graham began by discussing simulation which is essentially a tool of SRE.

> *"Today we tend to go on for years, with tremendous investments, to find that the system, which was not well understood to start with, does not work as anticipated. We work like the Wright brothers built airplanes: build the whole thing, push it off the cliff, let it crash, and start over again. Simulation is a way to do trial and error experiments. If the system is simulated at each level of design, errors can be found, and the performance checked at an early stage".* (Naur, 1968)

This thinking represents a foreshadowing of reliability acceptance testing which lays out an approach to quantitatively understanding how much (or how little) testing is required (Musa 1987; Cusick 1993). Amusingly, this quote also is one that was often repeated in the industry and it is interesting to see that it has been with us from the beginning.
Another core method and practice of SRE is the Operational Profile (Musa, 1993). The seeds of the Operational Profile which calls for understanding the feature set of an applications along with the probalistic level of being invoked. These drives testing which drives an understanding of reliability. At the NATO Conference, Perlis (Naur, 1968), stated that "… completeness means that the system must be capable of performing at least a 'basic' set of operations …" pointing to the future logic of the Operational Profile.

A final anecdote from the conference is worth mentioning as it relates both to perceptions of software reliability and to how we have all become accustomed to subpar software behavior. Smith discussed how:

> *"…if the users are convinced that if catastrophes occur the system will come up again shortly, and if the responses of the system are quick enough to allow them to recover from random errors quickly, then they are fairly comfortable with what is essentially an unreliable system."* (Naur, 1968)

This discussion found the author thinking of many conversations about how we have all learned to accept failure in our software applications. Essentially, we all have been conditioned to know that rebooting a machine is often going to be a quick fix. In reality, the real failure was in the design of the software and lack of reliability modeling prior to shipment. It is instructive how some of our modern apps for our mobile devices have become more reliable as they run in a highly restrictive computing environment. In any case, it is clear that the attendees of the NATO conference both foresaw many of the key aspects of SRE that would need to be solved and that many of the issues they lived with overall are still with us today or helped form our current state of practice in Software Engineering.

## V. Software Engineering and Measurement

Emerging from the NATO conference were various streams of research on software metrics and software measurement. Much of this was related to reducing software complexity meant to improve quality but also metrics to support estimation techniques. However, along with this research the foundations of the "engineering" side of Software Engineering began taking shape and this is where the Software Reliability Engineering methods were born.

To better understand the emergence of Software Reliability Engineering a look at the accepted definitions for Software Engineering is required. There are many definitions of software engineering and there has been some disagreement around what is meant by the term (Cusick, 2001). The following definition is instructive as it is both broadly accepted and other definitions tend to incorporate this one. Of note, the definition clearly mentions reliability, which is our primary concern in this paper.

> *"The establishment and use of sound engineering principles in order to obtain economically software that is reliable and works on real machines".* (Bauer, 1977)

This definition is useful as it mentions "engineering principles" and "reliable". From the point of view of engineering principles, it is expected that measurement would be a part of this approach and practice. Furthermore, reliability itself is only understood through the use of measurement. Thus, to define SRE we must first define what we mean by software measurement.

For measurement in general and reliability in particular the most important thing is that if we are doing some form of measurement, we are able to observe trends, set and track goals, and observe deviations in performance. Through measurement the efficiency and efficacy of new processes and tools can also be found. A standard statistical model for this was developed by Shewhart (Shewhart, 1931; Bolles, 2004) and is found in the standard statistical control chart in use across industries (see Figure 1). In this form of the control chart we monitor ongoing performance of the software application or other processes and then introduce new technology and observe the difference in effect. This is a key aspect of reliability engineering as well as general software measurement and process improvement methods.

A good metric is also constant in its application across systems. It is desirable for the metric to behave the same way and produce comparable information from software target to software target. Key metrics to collect by any means should include efficiency, cycle time, failures, product size, cost, and reliability. With software reliability it will be shown that it delivers a metric that is both consistent across systems and highly useful in measuring failures frequencies and operational effectiveness.

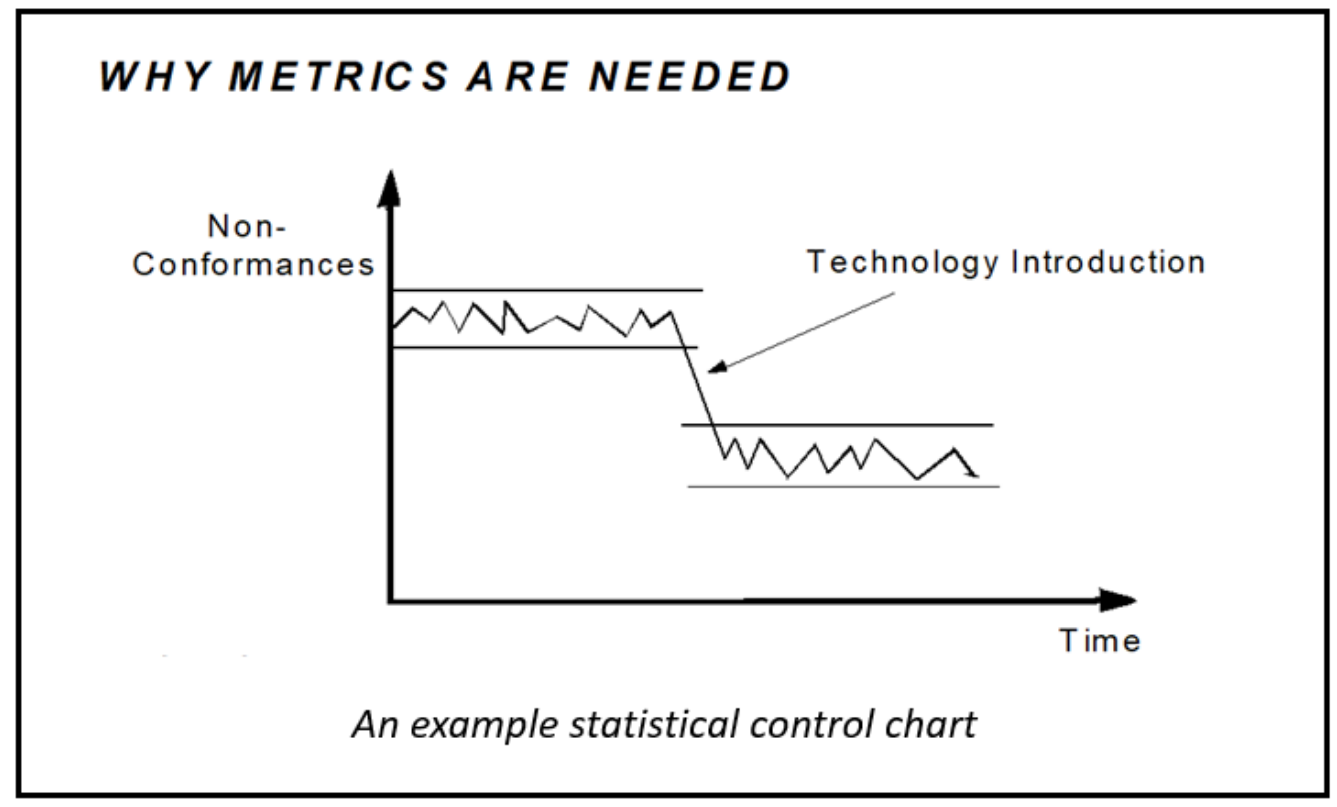

*Figure 1 – Metrics Application Demonstrated* (Glass, 1990)

Starting in the 1970s and improving in maturity through the 1990s, software metrics in support of Software Engineering included a wide array of specific metrics. These metrics underwent considerable research and industry usage with countless reports on their practical benefit. Some of the most popular and useful have been McCabe's Cyclomatic Complexity, Halsted's metrics, and various cost models such as Boehm's COCOMO (Cusick, 2013; Zuse, 2019). Other key metrics were the use of Lines of Code to conduct many types of sizing, costing, estimation, and quality analysis. Later, Albrecht developed Function Points which were further matured by Jones (Cusick, 2013). This metric represented a truly portable measurement across system types.

The development of this wide array of measurements took place in parallel to the development of SRE methods. For example, McCabe's impactful complexity paper was published in 1976 just 1 year after Musa's landmark paper on reliability models (McCabe, 1976). In many cases the development and use of these metrics were complimentary. For example, complexity has been shown to be linked to reliability. Additionally, researchers and practitioners met and shared results as these ideas formed at places like NASA's Software Engineering Laboratory's annual conference and many other venues (Cusick, 2018). Thus, it should be stated that SRE did not emerge in a vacuum but from within the Software Engineering community at large. In fact, the purpose of many software metrics was always to improve the quality and reliability of software, and it turns out that SRE was an excellent method to demonstrate if this had been achieved.

## VI. SOFTWARE RELIABILITY FUNDAMENTALS

Software measurements as noted above have various uses. However, as some software applications need to be built with a known level of reliability, thus a method to prove this was required. To fulfill this need a science and engineering practice has emerged over the last 50 years beginning at least in 1967 (Musa, 2006) which provides the tools, techniques, models, and methods to conduct software reliability engineering (SRE) successfully.

Software Reliability Engineering consists of processes and statistical methods used in predicting and tracking software reliability and related measures. Software reliability is generally defined as follows:

> *"The quantitative study of the operational behavior of software-based systems with respect to user requirements concerning reliability."* (Lyu, 1996)

or

> *"Software reliability is defined as the probability of failure-free software operation for a specified period of time in a specified environment."* (Walker, 1998)

Both definitions call for failure-free software. This software must also not only provide correct answers, but the software must also provide these correct results when the customer or user requests it to. Thus, the software must be reliable. This can be a very difficult requirement to fulfill and requires specialized skills and knowledge to do so predictably. When we consider systems such as the Space Shuttle Flight Control software, air-traffic control software, or real-time device control systems such as ABS (Anti-lock Brake System), you want the probability of successful operation of the software to be within the engineered specifications. This is where software reliability comes into its own. It is on such systems that a measure of how and when the software will fail becomes mission or at times life critical. Reliability engineering provides tools to answer this question and is a practice that has been deeply researched by the software engineering community and is widely practiced by leading software makers around the world not only for life-critical systems for form many other classes of applications as well.

In the beginning the field did borrow from traditional hardware reliability methods (Schneidewind, 2006) and in those environments, there is a traditional "bathtub curve" of reliability. In software this model does not apply. Instead we often see an operational reliability behavior as shown in Figure 2. In this view reliability increases during testing (or the failure rate declines). During operations spikes in the failure rate occur when change is introduced and finally the application enters obsolesce (Eusgeld, 2005). A major emphasis on SRE is in fact on how to determine the failure rates and related operational reliability as shown in this diagram.

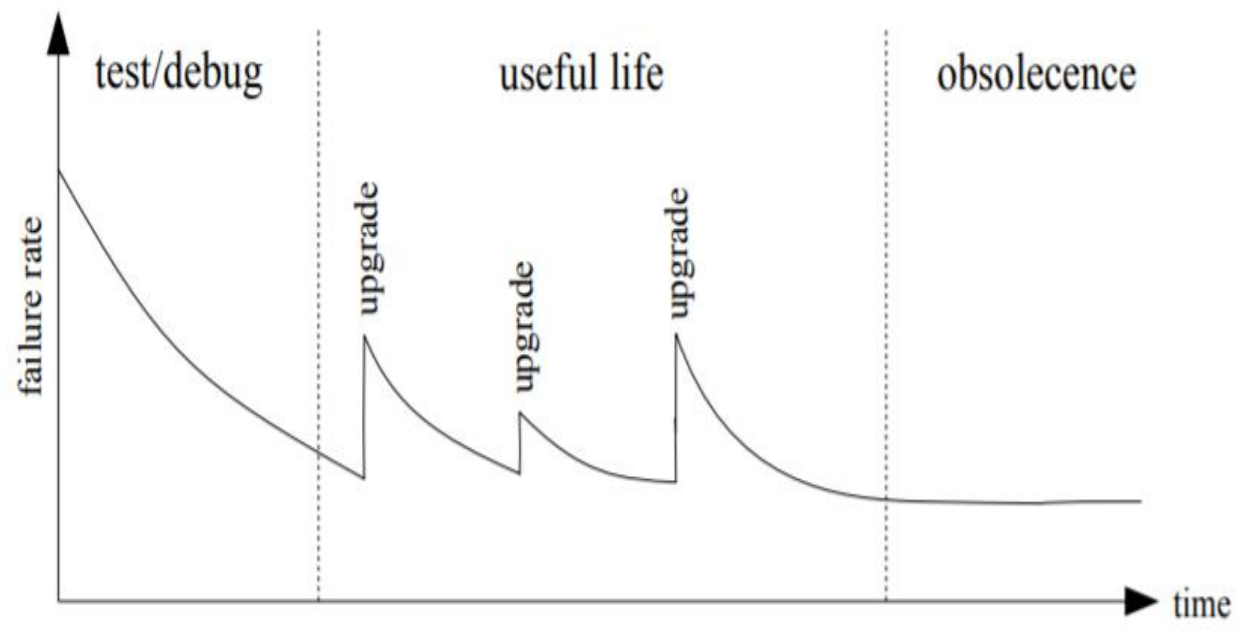

*Figure 2 – Idealized Software Reliability Model Over Time* (Eusgeld, 2005)

To understand the software reliability behavior of a given application it must be modelled and there are many models which now exist. In concept all of these models rely on the observation of failures over time. A common model is the Basic Execution Time Model of Software Reliability as presented by Musa (1987). This model provides a statistical methodology for measuring reliability and predicting reliability growth. The model allows for the calculation of reliability based on the non-normal distribution of software failures. The widely accepted equation below for reliability yields a single number which represents the probability of software executing without a failure for a given time period (Musa, 1987).

$$R(\tau) = exp(-\lambda\tau)$$

*where*

$R(\tau)$ = *reliability for time* $\tau$
$exp = e^x$
$\lambda$ = *failure intensity*
$\tau$ = *execution time*

The execution time above simply represents some measure of software usage such as CPU hours or clock cycles. As we shall see the creation of this approach was a novel milestone in the development of SRE. Failure Intensity represents the rate at which failures during operation of the software are encountered. Different usage patterns of any software often produce a variety of reliability levels. Operational Profiles quantify the usage pattern of the software by the intended users (Musa, 1993). Such a profile of utilization may be used to run tests on the system in order to track the reliability of the software as might be seen by the eventual users (Ackerman, 1989). This acts to ensure accuracy of the reliability projections between testing and production use. Figure 3 below demonstrates the inverse relationship between failure intensity and reliability over time. Given additional operational time and the improvement (lowering) failure intensity the application's reliability as measure will rise asymptotically.

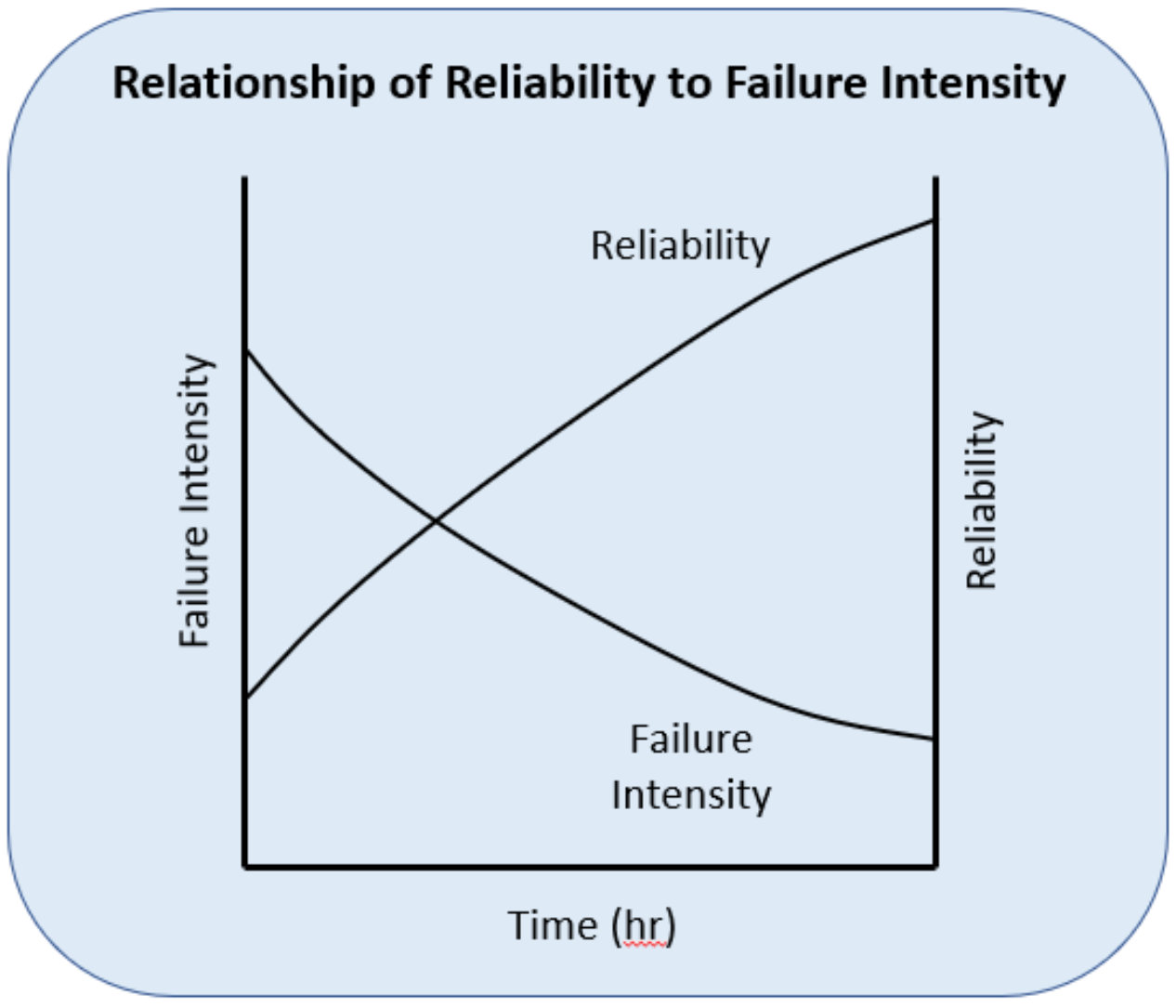

*Figure 3 – Reliability and Failure Intensity* (Musa, 1987)

Since its emergence as a discipline, the use of SRE has been widespread. SRE has been applied by numerous software producing companies and government agencies that have a demand for highly reliable systems. Organizations such as AT&T, Alcatel, Bellcore, CNES (France), ENEA (Italy), Ericsson Telecom, France Telecom, Hewlett Packard, Hitachi, IBM, Lockheed-Martin, Microsoft, Mitre, Motorola, NASA's Jet Propulsion Laboratory, NASA's Space Shuttle, Nortel, Raytheon, Saab Military Aircraft, Tandem Computers, the US Air Force, and the US Marine Corps have also used SRE effectively (Musa, 2006).

Finally, it should be noted that there is a rough split in the reliability world which can generally be seen as a "design time" reliability focus and an after the fact "reliability proving" focus. Much of SRE is focused on measuring reliability as the system emerges and prior to its deployment. It is not as focused on the design time aspects of reliability but can be used to confirm if such designs have met expectations. Various best practices in architecture, patterns, High Availability configurations, fault tolerance, and more all fall into the design time aspects that drive observed reliability as Djikstra indicated in 1968. Nevertheless, SRE remains a key practice in providing guidance to this design phase.

## VII. The Formation of Software Reliability

As presented in the review of the 1968 NATO Conference proceedings above, the topic of reliability was "top of mind" for many attendees and was considered a key attribute of software as well as a discipline area that was central to the newly established umbrella field of Software Engineering. Just prior to the conference in 1967, Hudson had developed the first reliability model for software (Hudson, 1967). This would mark the beginning of the field of SRE which would not be named as such for some time.

The formulation and interpretation of probabilistic reliability models in analysis and the utilization of these techniques for reliability design was detailed in a key book by Marty Shooman published in 1968 (Shooman, 1968). It was written as a text for college and industrial courses in reliability and for an engineering audience of a first-year graduate school or senior undergraduate level. This text provided an influential foundation for the further development of reliability methods as applied to software and building on Hudson's work (McLinn, 2010). This initial version was followed by a second edition with material on important new topics including fault-tolerant computers, software reliability, and risk analysis which was published over 20 years later (Shooman, 1990).

An interesting separation point in methods also takes place at this time. In 1968 simultaneous to the publications of Hudson's model, the NATO Conference, and Shooman's probability text, a fully revised RADC Reliability Handbook was issued (Mazzilli, 1968). RADC (Rome Air Development Center) had invested in documenting reliability methods for "systems" development as early as the mid-1950s. The first RADC Reliability Notebook was issued in 1958 and was the gold standard for reliability design for military grade systems. The newly written 1968 version contains not a single mention of the word "software" in its entire 361 pages. The word "application" is used often but is used to describe applying reliability methods to hardware and system solutions not as a modern reference to a "software application". This leads one to conclude that the Software Engineering domain had yet to make an impression in the systems world at this time, but it would do so shortly.

The focus and expansion on reliability theories and methods coincided with the emergence of large scale and life critical software applications. The early days of computing saw the development focused on batch-oriented data processing systems. At the time of the NATO conference entirely new classes of systems were emerging which required more predictable reliability and where software began to be the more complex and higher cost component to solutions (McLinn, 2010; Cusick, 2013). When software systems began to take center stage over hardware components in terms of cost and complexity, software reliability began to receive much deeper interest. Such methods and models began to emerge more fully as early as 1971 in what turned out to be cases of parallel and simultaneous invention:

> *"… in 1971 Jelinski and Moranda were developing software reliability models for the US Navy in San*

> *Diego for application to the Navy Tactical Data System (NTDS). This period was characterized by much doubt expressed by both academics and practitioners as to the validity of software reliability models. A frequent comment of contractors was "our software does not fail".* (Schneidewind, 2006)

However, in operation the software did fail, and this required a response from the system and software engineers on how to manage this tendency. Early work on these problems led to the exchange of research information at industry conferences such as one in April 1975 on software reliability in Los Angeles. This and other conferences brought reliability pioneers together. This conference also marked a key development in SRE as it brought together John Musa and Norm Schneidewind for the first time which initiated a long-standing collaboration. Norm Schneidewind started his research into software reliability in 1971 and John Musa began his work in this area in 1973. Their collaborations eventually influenced core developments in SRE and culminated first in an early paper by Musa published in 1975 in "IEEE Transactions on Software Engineering" (Musa, 1975). In this paper John was to help convert standalone models into a method of practice later called SRE. "This paper was to serve as a springboard for many researchers' efforts" according to Norm (Schneidewind, 2006). In fact, today this paper is widely cited by hundreds of researchers and practitioners on software reliability topics.

In conversation with John Musa, an SRE pioneer and leader in the establishment and spread of the field, he supported this same notion and noted that:

> *"... there was a lot of interaction between science and engineering and also contributions on the part of practitioners. I have always had the sense that practitioners and researchers mixed to an unusual degree and to an unusually international extent [at] ISSRE (International Symposium on SRE) ... a very innovative conference (fast abstracts, student papers, financial support for student attendance, "Anything You Want to Ask About SRE" panel, etc.)."* (Musa, 2006)

These early years of the emergence of the SRE field were well documented by Shooman's "History of Software Reliability" (1984) discussed below and mentioned by McLinn (2010). As the 1980s wore on these methods were further developed culminating in Musa's foundational book on "Software Reliability, Measurement, Prediction, and Application" (Musa, 1987). This book brought together all the theory, practice, and models required to effectively apply SRE as know at that time. In the following sections this timeline will be explored in more detail.

An interesting trend from the early years of SRE's development was the emergence of the field from several major government and especially military projects. From the NIKE Anti-Missile project to the US Navy NTDS project to the Space Shuttle, many of the major early developments in SRE found a basis in funding and application in the defense or space domain. Marianna Mazzucato of University College London who studies Innovation in the context of Public Value has noted that much of today's advanced technologies find at least some of their origins in publicly financed and directed programs (Mazzucato, 2013). It is clear from reviewing the early development of SRE that the field certainly emerged with a boost from various governments especially the US DoD.

## VIII. Early Reliability Treatments

Two major treatments emerged on software reliability in the late 1970s and early 1980s. The first is a book written by Glen Meyers and the second a detailed article by Marty Shooman. It is worth reviewing the contents of these two publications to provide further context and details on the maturation of the practice of software reliability and later coinage of the term SRE.

### *A. Meyers on Software Reliability*

Glen Meyers published his book "Software Reliability" in 1976 (Meyers, 1976). It is possible that this is the first book entitled as such and dedicated to the topic. In any case it is certainly an early text on computing and software reliability. Meyers was at the time a Research Staff Member at IBM's Systems Research Institute and a Lecturer in Computer Science at Polytechnic Institute of New York. The book focuses on many aspects of producing reliable software including both the "design time" issues and the "runtime" issues. In this book Meyers notes that in 1952 it was reported that some defects would escape detection for a long time even after release. This indicates an early understanding by authors and practitioners of the key objectives of designing and testing for reliability.

A favorite story of this author which is conveyed in the book is an early chapter entitled "Is the moon an enemy rocket?". In this story a missile detection system alerts its users of an incoming rocket which turns out to be the rising moon. This is the classical conundrum of needing to understand requirements fully in order to determine what is and what is not a failure. This pattern is one repeated often in Software Engineering and SRE. Meyers earns credit for documenting it so early.

Additionally, Meyers focused on complexity as a cause for unreliable software. This harkens back to the NATO conference nearly a decade earlier. He also discusses architecture and design methods - including error prediction, error tolerance, and fault correction. He even talks about how Human Factors play a role as usable design can impact operational reliability.

This book offers an early treatment of holistic testing approaches and methods very similar to our current understanding but also includes a view into the impact or cost to the user. Meyers rounds out his book by presenting the software reliability growth models (SRGM) extant at the time. This includes the reliability models from Jelinski & Moranda, Shooman, Schneidewind, and Musa. He concludes by reviewing various standard operational metrics such as MTTF (Mean Time To Failure). Reviewing this book again after a first read 25 years ago, it does serve as a tour de force of reliability in its day.

*B. Shooman on Software Reliability*

While the 1980s were taking off, software reliability techniques had become far more developed. When Marty Shooman published his comprehensive 1984 history of reliability, which remains relevant today (covering both hardware and software), the situation with Software Engineering had changed greatly since the NATO conference in 1968 only 16 years earlier. While many of the same problems might continue to be encountered by developers, the tools and methods had matured and the practice of SRE had begun to spread widely in industry and government. Furthermore, a sea change had occurred in technology where it was now software driving systems development not hardware alone. However, a review of Shooman's history of reliability from this time is in order to cement the view of those early decades of methodological advances (Shooman, 1984).

Marty Shooman graduated from MIT and Polytechnic Institute of New York with degrees in Electrical Engineering and is a Fellow of the IEEE. He worked for companies including GE, RCA, and Grumman. However, he spent the bulk of his career as a Professor at the Polytechnic Institutes of Brooklyn and New York and received numerous awards for contributions to the field. His text "Probabilistic Reliability: An Engineering Approach", previously mentioned (Shooman, 1968), was a notable addition to the understanding of reliability and is a key marker in the establishment of applied software reliability by providing the statistical foundations for many future modeling methods.

In his 1984 paper "*Software Reliability: A Historical Perspective*", which appeared in "IEEE Transactions on Reliability" (Shooman, 1984), he covered the development of reliability engineering up to that time. It is in this paper that he frames the application of reliability engineering techniques to software during the preceding decade and a half. He states early in the paper that:

> *"The basic problem in the software area is that the complexity of the tasks which software must perform has grown faster than the technology for designing, testing, and managing software development."* (Shooman, 1984)

Shooman goes on to define software reliability in much the same terms as others had and is considered the standard definition today. Essentially his definition states that "… software reliability is the probability that a given software system operates for some time period without software error …". Shooman also confirms that "by the late 1960s software designers and theoreticians had begun to think a lot about reliable software and were groping with the concept of software reliability with little results". This adds credence to the timeline of the beginnings of software reliability methods being at approximately 1968.

Shooman introduces a useful representation of the components of system reliability below:

$$R_{SY} = R_S \, x \, R_H \, x \, R_O$$

*where:*

*R = Reliability*
*SY = System*
*S = Software*
*H = Hardware*
*O = Operators/Operations*

The relationship of reliability for each of these components then yields System (SY) reliability. He argues that software has taken on a significant force in determining $R_{SY.}$ His 1984 treatment of recent reliability history is tilted towards understanding software reliability.

According to Shooman one of the earliest sources of practical software reliability applications and data came from J. A. Harr of Bell Laboratories, one of the architects of the Number 1 ESS (Electronic Switching System). The design requirement for the original ESS was for no more than two hours of overall system downtime (both hardware and software) over 40 years. This forced the application of reliability methods on the software of the ESS in order to ascertain the availability rating for the switch.

Shooman then details the emergence of the initial software reliability models confirming the citations presented above:

> *"[The] Markov birth-death model in 1967 [from Hudson], and the fitting of a growth function to cumulative error removal curves in the late 1960s [by Ellingson mark the first models]. Neither of the two System Development Corporation Memos [by Hudson & Ellingson] appeared in the open literature, and the work by Barlow & Scheuer apparently was not applied to software development. [Thus] the earliest [published] software reliability models were independently developed by Jelinski & Moranda [jointly], and Shooman [independently]. Both were first circulated internally and then published in the open literature in 1971."* (Shooman, 1984)

In retrospect the credit does seem to sit with Hudson for the first reliability model, however, as his work was not published externally until later it is Jelinsky, Moranda, and Shooman who earn the bragging rights for the first well understood models.

Shooman focused on the lack of available data related to software failures and reliability in operations as a limit to the predictive ability of reliability growth models. The earliest data set was provided by John Musa who collected error and test data for 20 software projects at Bell Laboratories in the early 1970s (Musa, 2006). This data had an average deviation of 24% between test environments and production environments. Shooman's own research with Miyamoto published separate data findings with a variance range from 6% to 19% across a smaller number of projects (Shooman, 1984). The broad industry database Shooman advocated for did not necessarily materialize but reliability models did improve in predictive accuracy with better algorithms and datasets and published case studies partially substituted for database of findings.

Shooman also touches on availability in his history of reliability which he defines as "… the probability that a given software system is operating successfully, according to specifications at a given point in time." In fact, this is generally a more common metric in industrial settings than reliability. Customers often want to know what the "uptime" of the application is and not necessarily the reliability projection. Fortunately, availability can easily be computed from reliability (moving from a probability to a percentage) simply multiplying by 100 (Cusick, 2017):

$$R(t), \% = [(e^{\wedge}(-\lambda t)) * 100\%]$$

In developing availability SLAs (Service Level Agreements) the traditional approach is to construct an uptime target as a percent of the total planned operations window. To do so is yet another reason to deploy SRE. Without a reliability prediction, understanding projected availability is nearly impossible. In that case one can only report on observed availability and cannot engineer for a target level.

Shooman concludes his history by stating that the early research in SRE benefited from Rome Air Development Center (RADC). Shooman notes that this "program established a sizeable research program in the areas of software measurement, reliability, and engineering." He concludes by stating that SRE was viewed as an established practice in 1984 and that new research efforts were continuing such as the initiative kicked off in 1984 at the Naval Air Systems Command to investigate problems of system reliability modeling for avionics and the interplay of hardware and software failures. It is from this point that we will now explore some first-person experience reports on the development of SRE and then pick up from 1984 and move to the present day.

## IX. First person Accounts - Methodology

This paper originally began as a thesis proposal. Parts of this paper were researched while the author was pursuing studies in History of Science at Polytechnic Institute of NYU from 2006 to 2008 under the advisement of Prof. Romualdas Sviedrys. During that research effort two first person interviews were conducted to help prove out the data collection approach for the project, yet the results of those interviews were never fully published. These interviews were with the late Norm Schneidewind and the late John Musa who were both contributors to the formation of the field, its models, and its practice and who the author was fortunate to have known professionally. The interview with Dr. Schneidewind has not previously been published. Parts of the interview with John were published posthumously as a tribute to his contributions to the field (Cusick, 2009). That publication is sampled here and is expanded on with the inclusion of additional and previously unpublished parts of the interviews from the original notes.

One motivation for developing this paper was to finally get those interviews in print together with the surrounding context of SRE's history as was originally planned. The interviews with Norm and John took place separately with the author in 2006 over the course of several days by email and telephone (Shneidewind, 2006; Musa, 2006). Both John and Norm had an opportunity to review the transcripts of these interviews at the conclusion of the process.

These interviews were scripted covering approximately 30 questions and the interview process also contained some ad hoc questions and dialogue as needed during the conversation's development to clarify answers or pursue interesting leads. The questions were organized into the following categories:

1. Biographical Information
2. Historical Influences on SRE
3. The Science of SRE
4. The Engineering of SRE
5. Future of SRE

The intention of the interviews was to start with the scripted questions and then to follow the threads of the conversations exploring the roots of this field with people who were there at the beginning. The questions were sent to the participants in advance electronically so that they might prepare for the interview and even respond in writing should that be convenient.

## X. Interview with Norm Schneidewind (2006)

### *A. Professional Biography*

Dr. Norman F. Schneidewind (1928–2015) was Professor Emeritus of Information Sciences in the Department of Information Sciences and the Software Engineering Group at the Naval Postgraduate School and a Fellow of the IEEE. He was active in research and publishing software reliability and metrics.

Norm was the developer of the Schneidewind software reliability model used by NASA to assist in the prediction of software reliability of the Space Shuttle and by the Naval Surface Warfare Center for software reliability prediction on various systems. This model was later recommended by the American National Standards Institute and the American Institute of Aeronautics and Astronautics (AIAA) Recommended Practice for Software Reliability.

### *B. Interview with Norm*

**Q: Norm, can you provide some background on your education and how you started work in reliability?**

My grandmother got me started on reading books about engineering. I graduated from UC Berkeley with a PhD in Electrical Engineering. It was a seminal experience which was very tough, challenging, and shaped my life. I began my professional career at Hughes Aircraft Co, working on airborne computers. My early research was in operations models. It was in 1971 that I began work on software reliability as the Chairman of my Department at the Naval Post Graduate School told me about a need by the Navy to conduct research on software quality down in San Diego.

**Q: Can you describe the earliest years when the software reliability field began to jell?**

It would be helpful to distinguish between "reliability" and "Software Reliability Engineering". The former preceded the later by several years. I think software reliability was discussed as a concept at the NATO conference in 1969. I first heard about it in 1971. However, SRE was not formed as a term or a practice until 1975. The first people to work on software reliability were Jelinski, Moranda, Arnold Goodman, Tom Brereton, and myself (Norm). We were all working in parallel around the same time in 1971.

**Q: What were the assumptions and goals behind SRE in the beginning?**

We assumed software failures could be modeled by a stochastic process and that we could improve software reliability. There was some dependence on hardware reliability for basic concepts, but we departed from these methods based on decreasing failure rate for software versus constant rate for hardware. To develop models, we compared predictions with actual reliability to understand the accuracy of the models themselves as well as the achievement of the reliability objective for the application under test. These comparisons of prediction accuracy used MSE (Mean Square Error), residual error, and chi square. The tools SMERFS and CASRE later built in these types of capabilities to a degree.

**Q: What new concepts were required to be developed to form the practice, e.g., definitions of faults vs. failures, execution time?**

The ability to model software reliability with probability models required invention. Many experts believe that software fails deterministically because there is an error made by humans that cause faults in the code that lead to failures in operation. Einstein said God does not play dice! Yes, if we knew everything about everything, we would not need probability. However, the input space and program space of most software is so large that failure occurrence can be considered a random process.

**Q: What were the biggest challenges you faced in realizing success with SRE in the beginning?**

Gaining acceptance: contractors said their software did not fail!

**Q: As SRE was first applied what were the problems encountered? Did any issue arise that required revalidation of the methods?**

Of course, there are always changes in research approach, as we learn more. For example, I learned that risk should be included in reliability assessment.

**Q: What were the first critical steps to make SRE a field of practice for non-theorists to apply?**

The original AIAA Recommended Practice on Software Reliability was important. Also, Lyu's SRE Handbook and the latest draft of recommended practice: IEEE/AIAA P1633™/Draft 5.7 played a role. The Draft Recommended Practice for Software Reliability Prediction, prepared by the Software Reliability Engineering Working Group of the Definitions and Standards Committee of the Reliability Society, April 2007, is also important. Also, the ongoing ISSRE conference which shares information within the reliability community, the standards groups, other publications, and personal communication.

**Q: What areas of the field were developed by practitioners and then adopted within the practice?**

Applied models and reported results have been important. This has built up a useful body of knowledge. Also, standards have been developed by virtue of demonstrating validity in real world applications first.

**Q: What are the remaining problems to be solved in SRE?**

Models should predict accurately early in the life cycle with requirements visa vie data. Very difficult to do!

### *C. An Additional Note from Norm*

An important historical note is that in 1971 Jelinski and Moranda were developing software reliability models for the US Navy in San Diego for application to the Navy Tactical Data System (NTDS). This period was characterized by much doubt expressed by both academics and practitioners as to the validity of software reliability models. A frequent comment of contractors was "our software does not fail". In addition, some engineers stated, "there is no such thing as software reliability because it does not wear out"! Actually, as I learned from sailors in 1971, when working on software reliability for NTDS, software becomes "tired" when subject to overload from tracking many targets. Thus, in a sense, it does "wear out".[2]

## XI. Interview with John Musa (2006)

### *A. Professional Biography*

John Musa (1933–2009) contributed broadly to the field of Software Engineering, especially in the area of Software

[2] Author's note – This behavior might be better classified as a performance degradation under load.

Reliability Engineering (SRE). John was a Naval officer after finishing college and then joined Bell Laboratories where he worked on the Nike Zeus anti-missile system project. After becoming a Bell Labs Supervisor John developed several novel concepts and methods in software reliability and later coined the phrase Software Reliability Engineering. John was a prolific author, active in teaching reliability and after retiring from AT&T taught and consulted in the industry on reliability.

*B. Highlights of Tribute Article (2009)*

From the interviews John held with the this author in 2006 a tribute to John and his contributions to the field was eventually published in IEEE Computing in 2009 (Cusick, 2009). In support of the history described earlier in this paper it is instructive to quote some of these published discussions with John again here in a focused format:

> *"G.R. Hudson (Hudson, 1967) probably developed the first model around 1967. Zygmund Jelinski and Paul Moranda introduced a model in 1971. Norm Schneidewind and Marty Shooman also developed models and applied them to previously collected data. Bev Littlewood had his own models; he used a Bayesian approach. I was (as far as I know) the first to develop a model and apply it in real time to manage an industrial project. In effect, we developed [models] and a practice at the same time."* (Cusick, 2009)

In terms of John's own model development perhaps the most foundational contribution was the use of calendar time, as John stated: "I developed a model that borrowed from ideas of Shooman and Jelinski and Moranda but was based on execution time instead of calendar time*"*. From John's perspective "…most of the theory was developed from 1971 to 1983, and the practice evolved after that." (Cusick, 2009).

Aside from John's Musa-Okamoto SRGM (a standard in the field), the innovation of using calendar time published in his "watershed" 1975 paper (Musa, 1975), and his outreach in promoting SRE, perhaps the most impactful and practical method he developed and introduced was the Operational Profile. This has been mentioned earlier, however, in John's words:

> *"The 1993 IEEE Software article [which I wrote] was the first comprehensive treatment of the subject that was published aside from the book. The practice was understood years before it was formalized and published except at a high level."* (Cusick, 2009)

*C. Intro to New Musa Interview Content*

The author was in frequent contact with John since meeting at Bell Labs in 1994 and then up until his passing. He participated in the 2006 interview and provided copies of numerous original documents related to his early work on SRE at the time of the interviews. Due to space limitations in the 2009 tribute article not all of that material could be included. What follows in the next section are some additional insights into the development of the field and John's role in this as well as the contributions of others as mentioned by John.

*D. Previously Unpublished Musa Interview Content*

*"I [John] worked in systems engineering, data smoothing, and simulation on the Nike-Zeus anti-missile project for about 15 years. I moved into a supervisor role over time.*

*In this role I contracted with Harlan Mills to apply Structured Programming, a first at Bell Labs. This included the application of inspections and programming correctness proofs which Watson Labs was pioneering. Exposure to these ideas helped motivate me to look for new software development approaches going forward.*

*At that time the state of the art in software engineering was basically guessing when the software would be ready[3]. We developed reliability models to predict the readiness date [and improve on this insufficient state-of-the-art].*

*Originally software reliability work was very model oriented. When I started looking at the problem, I took a much more general approach. I wanted something simple and practical to figure out the reliability objective and how to get there. I thought there were many people who were playing with math and statistics. This is good for a master's thesis, but it was not solving the real problems that projects face.*

*After leaving the Nike-Zeus project SRE became a sideline for some years. Once the practice began being used outside of Bell Labs by groups like the Nuclear Regulatory Agency and companies like HP, people started taking a more serious look at it. Eventually, in the Operations Technology Center of AT&T's Network Services Division under its VP Richard (Dick) Machol SRE began to be promoted AT&T wide.*

*In fact, there is an interesting anecdote about this. Dick was on a visit to HP in Silicon Valley and he saw signs all over the building promoting a talk by John Musa of AT&T on Software Reliability Engineering. At that time, he had not yet met me even though we both worked for AT&T. He began to think that if HP was inviting me in for technology transfer sessions on reliability methods, perhaps he and AT&T should be looking at this closer.*

*Of course, there was opposition to these methods due to the resistance to process changes, unwillingness to learn new techniques, changes of approaches, and the fact that people in general may push back [on] new things. Projects in trouble are more willing to change and so we did find willing users of the methods over time.*

*In general, at AT&T, we had a jump on other people in this area. We were the first to be working on real projects and practitioners were involved from the beginning. This allowed us to relate to other practitioners. They heard about it and started picking it up, focusing on practice at scale. We developed procedures and taught courses. It flowed pretty well. This might be a lesson in advancing other new technologies as well."*

[3] Refer back to the earlier NATO Conference quote on the Wright brothers by Graham which say nearly the same thing.

## XII. Published Histories 1990 to 2010

Following Shooman's major paper on reliability history and the flourishing of the reliability practice in the 1980s, culminated in Musa's book in 1987 and his paper on the Operational Profile in 1993, several authors developed and shared highlights of the field from 1990 to 2010. The first of these came from Bill Everett and colleagues in 1998. Around the same time Michael Lyu's definitive handbook on SRE appeared. Michael also published a future focused road map on SRE in 2007 which touched on recent progress in the field. We will review the key points of these writings to bring this history closer to the current day from Shooman's work in 1984.

### *A. SRE in 90s by Everett, et. al.*

Bill Everett, another Bell Labs alumnus, along with his colleagues Keene and Nikora provided an updated history of SRE focusing on work carried out in the 1990s (Everett, 1998). Samuel Keene is the author of a SRGM and Allen Nikora is an established reliability expert at JPL. This 1998 paper updates the state of the practice from Shooman's 1984 paper.

This 1998 paper again confirms that Hudson's work in 1967 was the first significant study of software reliability. They also note that in the period of the 1990s "…few organizations even measure[d] software reliability and some of those who [did] only measure it from a historical perspective." They mention that a pervasive view remained in many development communities that software cannot break[4]. Instead the developers continue to put the onus on the customer to be the final "arbitrator" as to what a failure is[5].

Furthermore, the authors state that at the time, for many developers one defect per KSLOC would be considered a relatively good latent defect level. As a result "…1,000 KSLOC of code, [would] contain 1,OOO latent defects at shipment." This meant that these latent defects would be left for the customer to find. This opens an opportunity for SRE practices to help reduce the failure rates of applications.

As the decade ended there were over a hundred reliability models that had been developed (Iannino, 1994). Allen Nikora argues in the 1998 paper that so many models indicated a dynamic field and was a positive sign. However, in a private conversation between Musa and this author during our 2006 interviews (Musa, 2006) John noted that many researchers continue to tweak established reliability models including those which are commonly used and well proven. Normally this was done to attain a limited new benefit and at the expense of breaking new ground or more fully supporting and advancing the practice. Thus, perhaps there are two ways of looking at this "dynamism".

Nevertheless, standardization of a core set of models was achieved in this time period as previously mentioned by Schneidewind and described by Everett in this paper. The AIAA (American Institute of Aeronautics and Astronautics) guidebook recommends the following models (AIAA, 1993):

- The Schneidewind Model.
- The Jelinski/Moranda Model.
- Musa/Okumoto Logarithmic Poisson Execution Time Model.
- The Littlewood/Verrall Model.

The paper also reviews some of the major efforts in promoting the use of SRE as well as the sequence of some of the significant technical developments in recent years further supporting the maturation of the field. These included the following major promotional developments for SRE:

- 1988 – Bell Labs developed and delivered courses in SRE to train engineers in applying the methods.
- 1990 – ISSRE (IEEE International Symposium on Software Reliability Engineering) was established allowing for the sharing and dissemination of ideas, techniques, and case studies across industry and academia.
- 1992 – The SRE Best Current Practice was formalized at AT&T. This rigorous process further codified the practice for use at AT&T and later allowed standard approaches to flow to the industry.
- 1993 – The introduction of the Operational Profile more broadly and in a fuller form via IEEE Software article by Musa.

The following developments in SRE tools were also noted by Everett and team as helping mature the field and now reflect key historical markers in SRE's development:

- 1977 – AT&T SRE Toolkit.
- 1983 – SMERFS (Statistical Modeling and Estimation of Software Reliability Functions)
- from Naval Surface Warfare Center.
- 1988 - SRMP (Software Reliability Modeling Program) by Reliability and Statistical Consultants, Ltd.
- 1992 - CASRE (Computer Aided Software Reliability Estimation).

### *B. Lyu SRE Handbook 1996*

Michael Lyu is an IEEE Fellow and Professor in the Computer Science and Engineering Department of the Chinese University of Hong Kong. Previously he worked at the Jet Propulsion Laboratory, Bellcore, and Bell Labs. He was a founder of ISSRE and editor of the Handbook on Software Reliability Engineering published in 1996.

The Handbook does not focus on the history of SRE per se but instead makes history itself by consolidating and advancing the methodologies and practice in the field. In fact, in the foreword to the book, renown Computer Scientist Alfred V. Aho says that the handbook is an "…important milestone in software reliability history…" (Lyu, 1996).

[4] See Norm Schneidewind's similar comments on this same observation some 40 years ago.

[5] Refer to the Meyer's anecdote of the "Moon as an Enemy Rocket" above.

In specific, the book provides definitive definitions, theories, methods, and approaches for SRE and features contributions from the world's leading reliability experts. The book presents all aspects of software reliability measurement and prediction. The book also covers "design time" issues such as "…product design, the development process, system architecture, the operational environment, and their effects on reliability". Benchmark case studies from AT&T, JPL, Bellcore, IBM, Nortel, NASA, Hitachi, ALCATEL, and other organizations are also included. Finally, emerging research methods including software metrics, testing schemes, fault-tolerant software, fault-tree analysis, process simulation, and neural networks are also covered (Lyu, 1996). This wide and deep coverage of all aspects of the field makes the book both useful and historic in the development of SRE.

Lyu confirms in the handbook that the Musa-Okumoto model is both the most widely adopted and "first to use the actual execution time" of software as expressed in computational units. The handbook also dives deep into the inventory of SRGMs and their mechanisms. An excellent survey and comparison of SRGMs was provided by Wood coincident with the publication of the Handbook (Wood, 1996) and more recently by Traore (2019). For demonstration purposes a sample (and simplified) Poisson model from the handbook can be shown as follows:

$$\lambda(\tau) = \mu'(\tau) = \alpha f_\alpha(\tau)$$

This formula represents in a summary manner the probability density function over time given a failure rate of α. Exploring SRGMs the handbook states that:

> "*…[commonly] model types mostly include finite failure models with failure intensity as exponential, binomial types with per-fault hazard rates, and Poisson models of several types. Also, Bayesian models have been developed*". (Lyu, 1996)

The book also provides extensive tutorials on methods and practices like the Operational Profile and extensive case studies. Thus, in the final analysis, the handbook was in fact a historical landmark in the solidification of the field and practice of SRE by collecting theory, models, practice, and case studies. This collection has proven to be a foundation piece in the field as it remains popular as a reference and has garnered over 2,700 citations[6].

Interestingly, at the same time as the handbook was being produced and the author was applying some of these methods, a colleague at Bell Labs who focused on software performance engineering made some comments about these SRGMs. He stated that using these models was like "trying to predict quality by looking in the rearview mirror." By this he meant that making software fault rate predictions from ongoing observed failures in test only provided a stochastic projection of what had happened in relation to execution time spent but did little to predict the future. Having used these models on real applications the author would have to say this is not exactly true at least once the first cycle is complete. After having calibrated test predictions to production further predictions become more and more accurate but perhaps never all knowing. But some healthy skepticism is always warranted.

### *C. A 2007 SRE Roadmap from Lyu*

Just over a decade following the SRE Handbook, Lyu provided a detailed look forward on where SRE was headed (Lyu, 2007). In doing so he also revisited the state of SRE while updating the forecast for the field. First, in a clear round-up of approaches to achieving reliability Lyu provides definitions for four methods to achieve reliable software across the lifecycle:

1. ***"Fault prevention****: to avoid, by construction, fault occurrences.*
2. ***Fault removal****: to detect, by verification and validation, the existence of faults and eliminate them.*
3. ***Fault tolerance****: to provide, by redundancy, service complying with the specification in spite of faults having occurred or occurring.*
4. ***Fault/failure forecasting****: to estimate, by evaluation, the presence of faults and the occurrences and consequences of failures. This has been the main focus of software reliability modeling"* (Lyu, 2007)

By laying out these approaches to achieving reliability Lyu succeeds in tying together much of the earlier writings we have discussed above. Starting with the NATO conference and moving to Meyer's discussion of reliability there was a focus on fault prevention from the beginning of the field. The same is true with fault removal where testing was discussed at the NATO conference and every stage of SRE development. Fault tolerance was considered in the most primitive models where mathematically adding components drives reliability probability down. And finally, failure forecasting is at the heart of SRE.

Michael also mentions that there are new methods which can influence reliability. These include new architecture types such as SOA as well as the use of Open-Source software which presumably may be of high reliability due to its wide testing and usage. This breakdown of paths to achieve quality are in fact all supported by SRE at one level or another.

Turning to reliability models again, Lyu notes that in 2007 there were 100's of SRGMs in existence and more are published on a continuing basis. However, of these models there are three primary kinds:

> *"1) the error seeding and tagging approach,*
> *2) the data domain approach, and*
> *3) the time domain approach, which [is] the most popular one."* (Lyu, 2007)

[6] Citation count provided by Google Scholar.

Explaining these models Lyu states that:

> "*The basic principle of time domain software reliability modeling is to perform curve fitting of observed time-based failure data by a pre-specified model formula, such that the model can be parameterized with statistical techniques.*" (Lyu, 1996)

Finally, Lyu notes that observation has shown that the adjusted Non-Homogeneous Poisson Processes (NHPP) are more accurate as they take into account additional predictive factors.

*D. Neural Net Reliability Models*

While the SRE Handbook devoted a chapter to Neural Net Reliability models these models have not always gained significant notice. However, starting in early 1990s and continuing to today advanced or intelligence-based approaches began being further researched including Neural Networks, Genetic Approaches, and Vector Machines (Ahmadluei, 2015). Research on these methods continue today in hopes of finding better predictive results. Based on such results the recurrent neural network-based approach is computationally feasible. It also seems to be potentially helpful in decreasing the cost of testing by accurately estimating the reliability of software. (Behera, 2018; Noekhah, 2018). It would seem reasonable to continue this line of investigation as it might provide great utility in the future.

## XIII. Recent Methods of Reliability Attainment

Looking back to the beginning of our story at the NATO Conference in 1968 we recall that all the participants pointed to the fact that testing alone was not the way to deliver reliable software. It was clearly stated that design-time steps to obtain clear requirements, effective architecture and design, and non-complex coding styles combined to lead to highly reliable solutions at run-time. A wide array of methods was developed in the 1970s and 1980s to pursue this track (Cusick, 2013). Early development methods like structured analysis and design, defensive programming, and the afore mentioned metrics such as complexity all strove to improve quality and as a result improve reliability. We now attempt to bring this history up to 2018 from that last major review in 2007 by Lyu.

*A. Process Models and Object Technology*

From the 1990s forward much effort was placed on maturity models as a framework for process improvement to achieve better software quality (Humphreys, 1989). In parallel, innovations in Object Technologies including the emergence of Design Patterns led to much more reusable and containerized software design which brought inherent reliability improvements (Cusick, 1999). Today's applications are almost universally built with some degree of Object Technology especially at the language level. From C++ to C# to Java to Javascript, and many others, the built-in Object Orientation providing encapsulation, inheritance, and component design has improved reliability at design-time and run-time from the 1990s through the current day.

*B. Emergence of Agile and DevOps*

Eventually, the structured, plan-driven maturity approaches requiring significant process documentation mostly gave way to Agile Methods which generally continued to use Object Technologies. Among the key philosophies behind Agile is that "working software is the primary measure of progress" (Beck, 2001). Again, these methods harken back to the incremental concepts mentioned in 1968 but now the approaches have become more formalized.

While Agile does not necessarily endorse or prescribe a quantitative means to know if the software is actually working correctly such as with SRE methods it is assumed that solutions will meet requirements and deliver value (Beck, 2001). Some researchers and practitioners have investigated the application of traditional SRGMs and SRE to Agile development teams. For example, Rawat proposed an SRGM for software under Agile development using both an NHPP and using the Musa model (Rawat, 2017) and were able to demonstrate that the Musa model worked in an Agile environment.

> *"Two types of faults, i.e., permanent and transient, [were] treated independently for each release. [The] comparison of the reliability of the two SRGMs indicates that the Musa model based SRGM yields better reliability results than [a] NHPP based SRGM. The capability of the Musa model to [react to] substantial changes in software over time as faults are observed makes it perform better."* (Rawat, 2017)

In addition, there has been a mind shift in how software is released emanating from the Agile approach. Today, large web-based providers often launch defined beta trials using their Agile processes which millions of users are eager to try despite any potential quality or reliability issues. These users act as unpaid testers eager to see what is new and provide feedback to the development teams. This allows for a rapid evolution of the software and improvements to reliability. This approach will not necessarily work in a life-critical avionics application but has greatly influenced modern commercial development.

This constant high rate of release to millions of users has also been supported and enabled by the rise of DevOps. One salient aspect of DevOps as it relates to reliability is CI/CD (Continuous Integration/Continuous Development). This method of build, deploy, and release relies on end-to-end automation including automated in-process testing. If any component breaks the build or fails a test it is shunted out of the deployment which is managed via feature flags. This keeps the main release flow moving and assures that working code is delivered every time.

A real-world example of this is the Facebook platform. Recently the chief build engineer at Facebook published an experience report on how the company designed its Continuous Deployment methods and environment to meet significant global demand from its development base (Rossi, 2017). The process has now been scaled from supporting hundreds of changes per day to supporting over 50,000 builds a day in its mobile environment. Such high volumes of releases were unthinkable at the time of the NATO Conference but are state-of-the-art today and help drive applications to higher levels of reliability even if SRE is not applied.

*C. Web Software Reliability*

For many modern applications a major platform of choice for deliver is in fact the World Wide Web. For Internet based applications or web applications (such as Facebook mentioned above) there is no inherent incompatibility with the use of SRE methods and such use of SRE on web applications is widely documented. However, many of them do follow Agile and/or DevOps methods. Albeanu (2019), as an example, demonstrated applicability of SRE to modern web software engineering environments, architectures, and methods covering multiple SRGMs. Building on the use of Object Technologies, Agile development, and other methods such as UML notation, navigation-based templates, hypermedia modeling based on object thinking, and documented models, Albeanu demonstrated how these approaches can help to further improve reliability on web platforms.

Anecdotally the author has observed that the Operational Profile or a similar technique is seen as very popular in the development community today whether on Agile teams or web development teams. This method has been used in feature prioritization on Agile development teams as well as performance analysis, and test selection.

*D. Mobile Development*

Mobile device application development is currently a significant market and the use of reliability methods for those applications while required are still in an early stage of validation. In fact, for many developers the SRE methodologies are either not well known or are not generally adopted.

An early treatment of SRE as applied to mobile devices actually comes from one of the founders of SRE methods, Norm Schneidewind. Norm developed a tutorial on recommended practices for reliability as applied to mobile devices (Schneidewind, 2008). This approach was taken as a starting point the IEEE/AIAA Reliability Recommended Practice but extended it to mobile devices.

As part of Norm's studies, he reported that for Symbian OS-based smart phones failure data were collected from 25 phones (in Italy and the US) over a period of 14 months. Key findings indicated that: "(i) the majority of kernel exceptions are due to memory access violation errors (56%) and heap management problems (18%)". Such studies began putting some quantification around the source of failures in these environments.

Building on this, Capretz (2013) conducted a detailed study into the application of SRE on mobile devices. The first step was to classify the types of failures typically found. Capretz found that failure causes in mobile environments does not look all that different than traditional computing failure issues and include those listed in Table 1 below.

| Wireless Failure Types | |
|---|---|
| Code | 3rd Party Software |
| Interfaces | Wireless Network |
| Hardware | Mobile Database |
| Interaction | OS Version |
| Data Input | Software Upgrades |

*Table 1 – Common mobile failure categories*

Next, the study selected three of the most used models: the NHPP, Musa-Basic, and the Musa-Okumoto models. Testing was done on two common iPhone applications: Skype and Vtok. Capretz found that none of the selected SRGMs was able to account for the failure data satisfactorily (Capretz, 2013). The research indicated that the models did not adapt well to the bursty rate of failures, and this may mean that new types of SRGMs might be called for to provide utility to the mobile environment.

Another useful study in this area comes from Meskini (2013). The main findings of this work are:

> *1. "The smartphone applications and their failure rates show distinctive features that differ from those of desktop/laptops ... requiring adjustments to the usual SRGMs.*
> *2. A reliability model suited to assess and predict the reliability of smartphone applications and their operational conditions is still needed.*
> *3. No one single distribution can account for all the failure data of an application through all of its releases. Nevertheless, the Gamma distribution ... and the S-shaped distribution, are more frequently suited to model the failure data."* (Meskini, 2013)

In summation, mobile development can be seen as similar to traditional environments in a number of ways, but it also calls for some new SRE methods which are still emerging.

*E. Service Management Advances*

A final area that has contributed to improvements in reliability achievement is the use of formal IT Service Management processes like the ITIL (Information Technology Infrastructure Library) framework (Bon, 2005). As noted by Shooman earlier, system reliability is dependent upon operational factors as well as software and hardware factors:

$$R_{SY} = R_S \, x \, R_H \, x \, R_O$$

The ITIL framework provides best practices across a wide range of operational functions and capabilities. In many organizations these processes are not only defined but automated helping to reduce error rates in conducting routine IT operations such as instantiating resources and assets or triggering routine maintenance events all of which can impact reliability (Cusick, 2017).

## XIV. Conclusions

Our review of Software Reliability Engineering history has taken us from the first reliability models and the NATO Conference of 1968 to Agile Methods and smartphones. In between we have seen how the earliest models were developed and the methods spread into the industry. We have also witnessed the leaders in the field and their lasting contributions. This historical review has shown the process of evolution in the field. Finally, we have seen where more work remains to be done to engender relevant support for the application of SRE in new environments.

While SRE was always a niche practice within Software Engineering most often applied by large-scale or life critical systems development many other companies also applied SRE. Two bibliographies are included following the references: 1) key historical papers; and 2) experience reports. In terms of the state of the practice today, notable companies who remain active in SRE include NASA, Bell Labs, IBM, Google, VMWare, Microsoft, Cisco, and others as shown by their recent participation in ISSSRE (ISSRE, 2018). In addition, considerable research continues to be pursued predominantly by academic institutions publishing in this area (ISSRE, 2018).

Unfortunately, some new publications may not contextualize the 50 years of history just presented. Instead when they ask the question "What's new in reliability", they look at combining McCabe's 1976 Cyclomatic Complexity metric for example with traditional SRGMs as in Luis Roca (2018). Naturally, this is a question that was answered fully years ago. Thus, without understanding the history of Software Engineering, software measurement, and SRE, little in the way of new methods or understanding can be created. It seems that some new research is simply revalidating decades of prior art without necessarily acknowledging this context or adding significant new progress to the field. It is hoped this paper can help remind such researchers where to look for known practice and methods and where the gaps remain.

In conclusion, the first 50 years of Software Reliability Engineering has brought tremendous progress in concepts, theory, models, methods, applications, and case studies. More than this, applications of SRE have led to more reliable and thus more trustworthy and safer software applications across a wide range of industries and platforms. It is not hyperbole to state the SRE's first 50 years has made a significant contribution to the usefulness of systems in our modern Information Age and it is likely to continue such contributions in the future ongoing digital transformations across our industries and affecting our lives.

## XV. Acknowledgements

The introduction of the author to Software Reliability Engineering was on the teleconferencing projects at AT&T. Frank Ackerman and Bill Everett from Bell Labs assisted in my initial work in this area. Not long after I had the pleasure to meet with and work with John Musa and other key people working on reliability at AT&T including Michael Lyu, Elaine Weyuker, and Karl Rauscher. I appreciate all they taught me. Later I was invited to present my early case study in this area at NASA's Software Engineering Laboratory and I met such people as Vic Basili, Norm Schneidewind, and other leaders in the field. To them and many more I am indebted. Finally, to Professor of History of Science Romualdas Sviedrys at NYU Polytechnic Institute I appreciate the support provided in the early stages of this research project.

## XVI. About the Author

James Cusick is Director, ITSM Process Management with a global Information Services firm. James is an IT leader with over 30 years of experience in Software Engineering, IT Operations, Information Security, and Project Management. James has held leadership roles with companies including Dell, Lucent's Bell Labs, and AT&T Laboratories. James applied SRE methods at AT&T and other companies and has written accounts relating to these experiences. James was also an Adjunct Professor at Columbia University where he taught Software Engineering. He has published widely in the above areas including two recent books on Software Engineering topics. James is a graduate of both the University of California at Santa Barbara and Columbia University. James is a Member of the IEEE and a PMP (Project Management Professional). Contact the author at j.cusick@computer.org.

## XVII. References

[1] Ackerman, F., and Musa, J. D., *"Quantifying Software Validation: When to Stop Testing?"*, **IEEE Software**, May 1989.

[2] Ahmadluei, S., Taxonomy of intelligence software reliability model, Journal of Advanced Computer Science & Technology, 4 (1) (2015) 60-67, 2015, www.sciencepubco.com/index.php/JACST.

[3] AIAA, "*Recommended Practice for Software Reliability: American National Standard (Ansi/Aiaa R-013-1992)*", American Institute of Aeronautics & Astronautics, February 1, 1993.

[4] AIAA, "Software Reliability Estimation and Prediction Handbook", 1992.

[5] Albeanu, G., et., al., Web Software Reliability Engineering (Vol.2) 2009, December, viewed on 1/12/2019 at https://www.researchgate.net/publication/228612739_Web_Software_Reliability_Engineering.

[6] Bauer, F. L., **Software Engineering: An Advanced Course**, Springer-Verlag, 1977.

[7] Beck, K., et. al., "*Manifesto for Agile Software Development*", **The Agile Alliance**, 2001, https://agilemanifesto.org/.

[8] Behera, Ranjan Kumar, et., al., Software Reliability Assessment using Machine Learning Technique, Computational Science and Its Applications – ICCSA 2018, July 2018.

[9] Bolles, D., ed., **A Guide to the Project Management Body of Knowledge**, 3rd ed., ANSI /PMI 99-001-2004, Project Management Institute, 2004.

[10] Bon, J. V., **Foundations of IT Service Management: based on ITIL**, Van Haren Publishing; 2nd edition, September 15, 2005.

[11] Capretz, Luiz Fernando, Meskini, Sonia, & Nassif, Ali Bou., "*Reliability Models Applied to Mobile Applications*", **Electrical and Computer Engineering Publications**, 2013.

[12] Card, R. Glass, D., **Measuring Sorfware Design Quality**, Prentice Hall, NJ, 1990.

[13] Chuck Rossi, "*Rapid release at massive scale"*, **Facebook Code Blog**, https://code.facebook.com/posts/270314900139291/rapid-release-at-massive-scale/, viewed 9/2/2017.

[14] Crosby, Philip B., **Quality is Free: The Art of Making Quality Certain**, McGraw-Hill Book Company, New York, 1979.

[15] Cusick, J., "*Software Engineering: Future or Oxymoron?*", **Software Quality Professional**, American Society for Quality, Volume 3, Number 3, June 2001.

[16] Cusick, J., "*Software Reliability Engineering for System Test and Production Support*", **Proceedings of the 4th International Conference on Applied Software Metrics**, Orlando, FL, November 1993.

[17] Cusick, J., & Cavolo, J., et. al., "*Managing a Corporate Object Technology Strategy*", **4th Conference on Object Technology Centers**, Baltimore, MD, June, 1999.

[18] Cusick, J., "*A Tribute to John Musa: In Memoriam*", **IEEE Computing**, January 2009.

[19] Cusick, J., "*Achieving and Managing Availability SLAs with ITIL Driven Processes, DevOps, and Workflow Tools*", **arXiv.org**, arXiv:1705.04906 [cs.SE], May 14, 2017.

[20] Cusick, J., **Durable Ideas in Software Engineering: Concepts, Methods, and Approaches from My Virtual Toolbox**, Bentham Science Publishers, 2013.

[21] Cusick, James, **Collected Papers on Software, Security, IT, and Technology: A 25 Year Retrospective**, Kindle Direct Publisher; 1st Edition, November 16, 2018.

[22] Cusick, James, *In Memoriam: John Musa*, Computing Now, **IEEE Computer**, January, 2009, https://www.computer.org/web/computingnow/musa.

[23] Eusgeld, Irene, "Software Reliability**", Conference on Dependability Metrics: Advanced Lectures Dagstuhl Seminar**, October 30 - November 1, 2005.

[24] Everett W., Keene S. Nikora, A. , "*Applying software reliability engineering in the 1990s*", **IEEE Transactions on Reliability**, Volume: 47 , Issue: 3 , Pages SP372 - SP378, Sep 1998.

[25] Goldstine, Herman, H., **The Computer from Pascal, to von Neumann**, Princeton University Press, 1972.

[26] Hudson, G., "*Program Errors as a Birth and Death Process*", **System Development Corporation Report SP-3011**, Santa Monica, CA, 1967

[27] Humphrey, Watts S., **Managing the Software Process**, Addison-Wesley Professional, January 1989.

[28] Iannino, A., **Software Reliability Theory in Encyclopedia of Software Engineering**, J. J. Marichiniak, ed., Wiley-Interscience, 1994.

[29] ISSRE, Memphis, TN, October, 2018, http://2018.issre.net/industry-pc

[30] Jelinski, Z. and Moranda, P., "*Software Reliability Research*", in W. Freiberger, ed., **Statistical Computer Performance Evaluation**, Academic Press, New York, NY, 1972, pp. 465-484.

[31] Jelinski, Z., Moranda, P. B., "*Software Reliability Research*," McDonnell Douglas Astronautics Co., Huntington Beach, CA., MADC Paper WD1808, pp 465-484, November 1971.

[32] Luis Roca, Jose, "*What's new about Software Reliability*", ResearchGate.net, October 2018, viewed 1/12/2019 at https://www.researchgate.net/publication/328638384_What%27s_new_about_Software_Reliability.

[33] Lyu, M., ed., **Handbook of Software Reliability Engineering**, IEEE Computer Society Press, Los Alamitos, CA, 1996.

[34] Mazzilli, F., et. al., "RADC Reliability Notebook", Volume I, Computer Applications Incorporated, Technical Report, RADC-TR-67-108, **Rome Air Development Center**, Air Force Systems Command, Griffiss Air Force Base, New York, November 1968.

[35] Mazzucato, Mariana, **The Entrepreneurial State: debunking public vs. private sector myths**, Anthem, 2013.

[36] McCabe, T., "*A Complexity Measure*". **IEEE Transactions on Software Engineering**, (4): 308–320, December 1976.

[37] McLinn, James, "*History Of Reliability Engineering*", **History of Reliability**, ASQ (American Society for Quality), April 28, 2010, viewed 1/1/2019. https://www.asqrd.org/home/history-of-reliability/.

[38] Meskini, Sonia, "*Reliability Models Applied to Smartphone Applications"*, A thesis submitted for the degree in Master of Engineering Science, The University of Western Ontario, 2013.

[39] Meyers, Glenford, J., Software Reliability: Principles and Practices, John Wiley & Sons, New York, 1976.

[40] Musa, J. D., "*A Theory of Software Reliability and its Application*," **IEEE Transactions on Software Engineering**, SE-1(3), Sept. 1975, pp 312- 327.

[41] Musa, J. D., *"Operational Profiles in Software-Reliability Engineering",* **IEEE Software**, March 1993.

[42] Musa, J. D., Iannino, A., and Okumoto, K., **Software Reliability: Measurements, Prediction, Application**, McGraw-Hill, 1987.

[43] Musa, John D., *"Conversations with the author",* via email and telephone, 2006.

[44] Musa, John D., "*User Experiences with SRE*", **Software Reliability Engineering and Testing Courses**, Updated March 28, 2002, Website viewed 2006.

[45] Naur, P., Randell, B., eds., **Software Engineering: Report on a conference sponsored by the NATO Science Committee**, Garmisch, Germany, October, 1968, viewed 1/19/2019, http://homepages.cs.ncl.ac.uk/brian.randell/NATO/nato1968.PDF

[46] Noekhah, Shirin, et. al., Predicting Software Reliability with a Novel Neural Network Approach, International

Conference of Reliable Information and Communication Technology, May 2018

[47] Pfleeger, Shari Lawrence, **Software Engineering, Theory and Practice**, Prentice Hall, Upper Saddle River, 1998.

[48] Rawat, S., et. al., "*Software Reliability Growth Modeling For Agile Software Development*", **Int. J. Appl. Math. Computer. Sci.**, 2017, Vol. 27, No. 4, 777–783.

[49] Saleh, J. H., & Marais, K., *Highlights from the early (and pre-) history of reliability engineering*, Reliability Engineering & Systems Safety, 2006, vol. 91, no2, pp. 249-256, viewed online on 12/28/2006: http://cat.inist.fr/?aModele=afficheN&cpsidt=17276535.

[50] Schneidewind, Norman, "*Tutorial on IEEE\AIAA Recommended Practice on Software Reliability Applied to Mobile Devices*", ISSRE,, 2008.

[51] Schniedewind, N., Conversations with the author via email covering Bio and Reliability History, 2006.

[52] Shewhart, W. A., **Economic control of quality of manufactured product**, D. Van Nostrand Company, New York, 1931.

[53] Shooman, Martin L., "*Software Reliability: A Historical Perspective*", **IEEE Transactions on Reliability**, vol R-33, pp. 48-55, April 1984.

[54] Shooman, Martin, L., Probabilistic Reliability: An Engineering Approach, Krieger Pub Co; Subsequent edition., August 1, 1990.

[55] Shooman, Martin. L., **Probabilistic reliability: An engineering approach**, McGraw-Hill, 1968.

[56] Tan, Cher Ming, "*Overview of Reliability Engineering*", **Theory and Practice of Quality and Reliability in Asia Industry**, C.M. Tan and T.N. Goh (eds.), 2017, DOI 10.1007/978-981-10-3290-5_2

[57] Traore, Issa, Software Quality Engineering, University of Victoria, viewed on 1/13/2019, http://www.ece.uvic.ca/~itraore/seng426-07/notes/qual07-8.pdf.

[58] van Vliet, **Software Engineering: Principles and Practice**, John Wiley & Sons, New York, 1993.

[59] Walker, Ellen, *Applying Software Reliability Engineering (SRE) to Build Reliable Software*, **Selected Topics in Assurance Related Technologies**, Dod Reliability Analysis Center, Rome, NY, Volume 5, Number 3, March, 1998 .

[60] Wood, Alan, Software Reliability Growth Models, Technical Report 96.1, Part Number: 130056, Tandem Computers, September 1996

[61] Zuse, Horst, History of Software Measurement, https://wwwbroy.in.tum.de/lehre/vorlesungen/vse/WS2004/zuseh_metrics_history.pdf, viewed on 1/13/2019.

## XVIII. Selected Historical References

Provided to the author via correspondence with Norm Schneidewind during the interview process, this selected bibliography lists many of the key references covering historical developments within the field of software reliability. This list covers the years 1972 through 1997.

1. Farr, W. H. and Smith, O. D., *Statistical Modeling and Estimation of Reliability Functions for Software (SMERFS) Users Guide*, NAVSWC TR-84-373, Revision 2, Naval Surface Warfare Center, Dahlgren, Virginia.

2. Farr, W. H., A Survey of Software Reliability Modeling and Estimating, Naval Surface Weapons Center, NSWC TR 82-171, September 1983, p. 4-88.

3. Goel, A. and Okumoto, K., "Time-Dependent Error-Detection Rate for Software Reliability and Other Performance Measures," *IEEE Transactions on Reliability*, Vol. R-28, No. 3, pp. 206-211.

4. Hecht, H. and Hecht. M., "Software Reliability in the System Context," *IEEE Transactions on Software Engineering*, January 1986.

5. Jelinski, Z. and Moranda, P., "Software Reliability Research," in W. Freiberger, ed., *Statistical Computer Performance Evaluation*, Academic Press, New York, NY, 1972, pp. 465-484.

6. Littlewood, B. "Software Reliability Model for Modular Program Structure," *IEEE Trans. on Reliability,* R-28, pp. 241-246, Aug. 1979.

7. Littlewood, B. and Verrall, J. L., (June 1974), "A Bayesian Reliability Model with a Stochastically Monotone Failure Rate," *IEEE Transactions on Reliability,* pp. 108-114.

8. Littlewood, B., Ghaly, A., and Chan, P. Y., "Tools for the Analysis of the Accuracy of Software Reliability Predictions", (Skwirzynski, J. K., Editor), *Software System Design Methods,* NATO ASI Series, F22, Springer-Verlag, Heidleberg, 1986, pp. 299-335.

9. Michael R. Lyu (Editor-in-Chief), Handbook of Software Reliability Engineering, Computer Society Press, Los Alamitos, CA and McGraw-Hill, New York, NY, 1995.

10. Musa, J. D., Iannino, A., and Okumoto, K., *Software Reliability: Measurement, Prediction, Application*. McGraw-Hill, New York, 1987.

11. Musa, J. D., Okumoto, K., "A Logarithmic Poisson Execution Time Model for Software Reliability Measurement," *Proceedings Seventh International Conference on Software Engineering*, Orlando, pp. 230-238.

12. Musa, J., "A Theory of Software Reliability and Its Application," *IEEE Trans. Software Eng.,* Vol. SE-1, No. 3, September 1975, pp 312-327.

13. Schneidewind, N. F. and Keller, T. M., "Applying Reliability Models to the Space Shuttle," *IEEE Software*, July 1992, pp. 28-33.

14. Schneidewind, N. F., "Software Reliability Model with Optimal Selection of Failure Data", IEEE Transactions on Software Engineering, Vol. 19, No. 11, November 1993, pp. 1095-1104.

15. Schneidewind, N. F., "Analysis of Error Processes in Computer Science," *Proceedings of the International Conference on Reliable Software*, IEEE Computer Society, 21-23 April 1975, pp. 337-346.

16. Schneidewind, Norman F., "Reliability Modeling for Safety Critical Software", IEEE Transactions on Reliability, Vol. 46, No.1, March 1997, pp.88-98.

17. Shooman, M. L. and Richeson, G., "Reliability of Shuttle Control Center Software," Proceedings Annual Reliability and Maintainability Symposium, January 1983, pp. 125-135.

18. Shooman, M. L. and Richeson, G., "Reliability of Shuttle Control Center Software," Proceedings Annual Reliability and Maintainability Symposium, January 1983, pp. 125-135.
19. Shooman, M. L., "Early Software Reliability Predictions," Software Reliability Newsletter, Technical Issues Contributions, IEEE Computer Society Committee on Software Engineering, Software Reliability Subcommittee, 9/11/90.
20. Shooman, M. L., "Structural Models for Software Reliability Prediction," *Second National Conf. on Software Reliability*, San Francisco, CA., October 1976.
21. Shooman, M. L., "Structural Models for Software Reliability Prediction," *Second National Conf. on Software Reliability*, San Francisco, CA., October 1976.
22. Shooman, M. L., *Probabilistic Reliability: An Engineering Approach*, McGraw-Hill Book Co., New York, NY, 1968, 2nd Edition, Krieger, Melbourne, FL, 1990.
23. Shooman, M. L., *Software Engineering: Design, Reliability, and Management*, McGraw-Hill Book Co., New York, NY, 1983.
24. Shooman, M. L., *Software Engineering: Design, Reliability, and Management*, McGraw-Hill Book Co., New York, NY, 1983.

## XIX. SRE User Experience Reports

As part of the interview process with John Musa he provided a list of user experience reports on the application of SRE. This list may be of use to those looking for not only the history of SRE but regarding how and where it was applied on real-world projects. In general research papers in SRE have been excluded from this list as they can be found in other bibliographic sources. In John's own words:

> *"I have also excluded articles and papers that analyze software reliability engineering data after the fact and those where no practitioners were involved. Thus, this list is intended for practitioners who want to understand what it is like to practice software reliability engineering in various settings."* (Musa, 2006).

1. Alam, M., W. Chen, W. Ehrlich, M. Engel, D. Kropfl, P. Verma. 1997. Assessing software reliability performance under highly critical but infrequent event occurrences. *Proceedings 8th International Symposium on Software Reliability Engineering*, Albuquerque, NM, November 1997, pp. 294-307.
2. Bennett, J., Denoncourt, M., and Healy, J. D. 1992. "Software Reliability Prediction for Telecommunication Systems," *Proc. 2nd Bellcore/Purdue Symposium on Issues in Software Reliability Estimation*, Oct. 1992, pp. 85-102.
3. Bentz, R. W. and C. D. Smith. 1996. Experience report for the software reliability program on a military system acquisition and development. *Proceedings 7th International Symposium on Software Reliability Engineering - Industrial Track*, White Plains NY, October 30-November 2, 1996, pp. 59-65.
4. Bergen, L. A. 1989. "A Practical Application of Software Reliability to a Large Scale Switching System," *IEEE International Workshop: Measurement of Quality During the Life Cycle*, Val David, Quebec, Canada, April 25-27, 1989.
5. Carman, D. W., Dolinsky, A. A., Lyu, M. R., and Yu, J. S. 1995. "Software Reliability Engineering Study of a Large-Scale Telecommunications Software System," *Proc. 1995 International Symposium on Software Reliability Engineering*, Toulouse, France, Oct. 1995, pp. 350-.
6. Carnes, P. 1997. "Software reliability in weapon systems." *Proceedings 8th International Symposium on Software Reliability Engineering: Case Studies*, Albuquerque, NM, November 1997, pp. 95-100.
7. Carnes, P. 1998. "Software reliability in weapon systems." *Proceedings 9th International Symposium on Software Reliability Engineering: Industrial Practices*, Paderborn, Germany, November 1998, pp.272-279.
8. Cramp, R., Vouk, M. A., and Jones, W. 1992. "On Operational Availability of a Large Software-Based Telecommunications System," *Proc. 3rd International Symposium on Software Reliability Engineering*, Research Triangle Park, NC, Oct. 7-10, 1992, pp. 358-366.
9. Cusick, J. and M. Fine. 1997. Guiding reengineering with the operational profile. *Proceedings 8th International Symposium on Software Reliability Engineering: Case Studies*, Albuquerque, NM, November 1997, pp. 15-25.
10. Derriennic, H. and G. Le Gall. 1995. Use of failure-intensity models in the software-validation phase for telecommunications. *IEEE Transactions on Reliability* 44(4):658-665.
11. Drake, H. D. and D. E. Wolting. 1987. Reliability theory applied to software testing. *Hewlett-Packard Journal* 38(4):35-39.
12. Ehrlich, W. K., Lee, K., and Molisani, R. H. 1990. "Applying Reliability Measurements: A Case Study," *IEEE Software*, March 1990.
13. Ehrlich, W. K., Stampfel, J. P., and Wu, J. R.
14. Elentukh, A. 1994. "System Reliability Policy at Motorola Codex," *Proc. 5th International Symposium on Software Reliability Engineering*, Monterey, CA, Nov. 6-9, 1994, pp. 289-293.
15. Hamilton, P. A. and Musa, J. D. 1978. "Measuring Reliability of Computation Center Software," *Proc. 3rd International Conference on Software Engineering*, pp. 29-36.
16. Iannino, A and Musa, J. D. 1991. "Software Reliability Engineering at AT&T," Apostolakis, G. (ed.) *Probability Safety Assessment and Management - Vol. 1*, Elsevier, New York.
17. Jenson, B. D. 1995. "A Software Reliability Engineering Success Story: AT&T's Definity (PBX," *Proc. 1995 International Symposium on Software Reliability Engineering*, Toulouse, France, Oct. 1995, pp. 338-343.
18. Jones, W. D. 1991. "Reliability Models for Very Large Software Systems in Industry," *Proc. 1991 International Symposium on Software Reliability Engineering*, Austin, TX, May 17-18, 1991, pp. 35-42.
19. Juhlin, B. D. 1992. "Implementing Operational Profiles to Measure System Reliability," *Proc. 3rd International Symposium on Software Reliability Engineering*, Research Triangle Park, NC, Oct. 7-10, 1992, pp. 286-295.

20. Juhlin, B. D. 1993. "Software Reliability Engineering in the System Test Process," *Proc. 10th International Conference on Testing Computer Software*, Washington, DC, June 14-17, 1993, pp. 97-115.

21. Keller, T. and N. Schneidewind. 1997. Successful application of software reliability engineering for the NASA space shuttle. *Proceedings 8th International Symposium on Software Reliability Engineering: Case Studies*, Albuquerque, NM, November 1997, pp. 71-82.

22. Kruger, G. A. 1989. Validation and further application of software reliability growth models. *Hewlett-Packard Journal* 40(4):75-79.

23. Lakey, Peter B. 1998. "How does any software organization proceed in incorporating SRE?" (Crusader self-propelled howitzer project) *Proc. 9th Annual SRE Workshop* 7/14-15/98, Ottawa, Ontario, Canada.

24. Lee, I. and R. K. Iyer. 1995. Software dependability in the Tandem GUARDIAN system. *IEEE Transactions on Software Engineering* 21(5):455-467.

25. Levendel, Y. 1989. "Defects and Reliability Analysis of Large Software Systems: Field Experience," *Proc. 19th IEEE International Symposium on Fault-Tolerant Computing*, Chicago, June 1989, pp. 238-244.

26. Levendel, Y. 1990. "Reliability Analysis of Large Software Systems: Defect Data Modeling," *IEEE Trans. Software Engineering*, vol. SE-16, no. 2, February 1990, pp. 141-152.

27. Mendiratta, Veena B. 1998. "Reliability Analysis of Clustered Architectures," Proc. 9th Annual *SRE Workshop* 7/14-15/98, Ottawa, Ontario, Canada.

28. Pemler, S. and Stahl, N. 1994. "An Automated Environment for Software Testing and Reliability Estimation," *Proc. 5th International Symposium on Software Reliability Engineering*, Monterey, CA, Nov. 6-9, 1994, pp. 312-317.

29. Rapp, B. 1990. "Application of Software Reliability Models in Medical Imaging Systems," Proc. *1990 International Symposium on Software Reliability Engineering*, Washington, DC, April 1990.

30. Sandfoss, R. V. and S. A. Meyer. 1997. Input requirements needed to produce an operational profile for a new telecommunications system. *Proceedings 8th International Symposium on Software Reliability Engineering: Case Studies*, Albuquerque, NM, November 1997, pp. 29-39.

31. Schneidewind, N. F. and Keller, T. W. 1992. "Application of Reliability Models to the Space Shuttle," *IEEE Software*, July 1992, pp. 28-33.

32. Teresinski, J. A. 1996. Software reliability: getting started. *Proceedings 7th International Symposium on Software Reliability Engineering - Industrial Track*, White Plains NY, October 30-November 2, 1996, pp. 39-47.

## XX. A Timeline of Software Reliability History

| YEAR | DEVELOPMENT OR EVENT |
|---|---|
| • **1816** | The term reliability coined by Coleridge. |
| • **1911** | Taylor's Scientific Management published. |
| • **1924** | Shewhart develops SPC charts at Bell Labs, later, in 1931 he publishes first statistical quality book. |
| • **1948** | IEEE Reliability Society formed. |
| • **1948** | DoD forms AGREE. |
| • **1964** | First software reliability data published by Harr of Bell Labs on 1ESS. |
| • **1967** | Hudson develops first software reliability model. *** ***Beginning of Software Reliability Methods*** |
| • **1968** | RADC new Reliability Handbook published. |
| • **1968** | NATO Conference on Software Engineering with section on reliability. |
| • **1968** | Shooman's text on Probablistic Reliability published. |
| • **1971** | Jelinski & Moranda publish a software reliability model. |
| • **1971** | Shooman publishes a software reliability model. |
| • **1975** | Musa IEEE article published on execution time model with project data. |
| • **1975** | SRE is coined as a term to describe Software Reliability Engineering. |
| • **1976** | Meyer's book on Software Reliability published. |
| • **1977** | Tool - AT&T SRE Toolkit. |
| • **1977** | Bauer offers definition for Software Engineering including reliability. |
| • **1983** | Tool - SMERFS from Naval Surface Warfare Center. |
| • **1984** | Shooman publishes history of reliability. |
| • **1987** | Musa, et. al., book on Software Reliability Methods published. |
| • **1989** | CMM is released to help drive software process improvement. |
| • **1988** | Bell Labs provides courses in SRE. |
| • **1988** | Tool - SRMP by Reliability and Statistical Consultants, Ltd. |
| • **1990** | ISSRE (IEEE Intl Symposium on SRE) established. |
| • **1992** | SRE Best Current Practice formalized at AT&T. |
| • **1992** | Tool - CASRE developed and released. |
| • **1993** | Operational Profile by Musa published in IEEE Software. |
| • **1993** | AIAA Reliability Guidebook released. |
| • **1996** | Lyu publishes SRE handbook. |
| • **1996** | Wood publishes SRGM survey. |
| • **1998** | Everett, et. al., provide SRE survey to date. |
| • **2001** | Agile Manifesto released. |
| • **2007** | Lyu publishes SRE roadmap. |
| • **2007** | Draft Reliability Practice published. |
| • **2008** | SRE tutorial for Mobile Development by Schneidwind. |
| • **2010** | Tool - SOFTREL. |
| • **2013** | SRE study for Mobile Development by Capretz. |
| • **2015** | Intelligence-based methods survey published by Ahmadluei. |
| • **2017** | SRGM usage for Agile presented. |
| • **2017** | CI/CD DevOps approach documented by Rossi to support throughput and reliability. |